\documentclass[aps,prl,reprint,nofootinbib,floatfix]{revtex4-2}
\usepackage[utf8]{inputenc}
\usepackage[T1]{fontenc}
\usepackage{amsmath,amssymb}
\usepackage{graphicx}
\usepackage{microtype}
\usepackage{bm}
\usepackage{xcolor}
\usepackage{hyperref}
\hypersetup{colorlinks=true,linkcolor=blue,citecolor=blue,urlcolor=blue}
\graphicspath{{./}{figures/}}
\begin{document}
\title{Atom-Resolved Machine Learning of Dielectric and Piezoelectric Response}
\author{Jinyu Liu}
\author{Yingwei Chen}
\author{Liyang Ma}
\author{Hongyu Yu}
\email[Corresponding author: ]{hongyuyu20@fudan.edu.cn}
\author{Hongjun Xiang}
\email[Corresponding author: ]{hxiang@fudan.edu.cn}
\affiliation{Key Laboratory of Computational Physical Sciences (Ministry of Education), Institute of Computational Physical Sciences, State Key Laboratory of Surface Physics, and Department of Physics, Fudan University, Shanghai, 200433, China}
\begin{abstract}
Predicting dielectric and piezoelectric responses in structurally complex materials requires large simulation cells, for which density-functional perturbation theory (DFPT) calculations scale as $\mathcal{O}(N^4)$ and become computationally prohibitive. However, machine-learning approaches have remained challenging, limited either by the large amounts of expensive DFPT data required for direct regression or by the cubic cost of reconstruction. We here propose a machine-learning framework that learns the ionic response as an atom-indexed field: DART (Direct Atom-Resolved response-Tensor learning) learns these fields directly and achieves high accuracy from small training sets, whereas LARS (Linear-scaling Atom-Resolved Response Solver) reconstructs them from learned microscopic ingredients through sparse linear solves and requires no DFPT labels for the ionic dielectric or piezoelectric tensors. Using DART, we extrapolate to 314 stackings of AlN/ScN superlattices absent from training and identify a high-response polar candidate, for which DFPT gives a laterally clamped piezoelectric strain coefficient $d_{33,f}=14.511$ pC/N and $k_t^2=20.46\%$, exceeding the ordered 1AlN/1ScN reference by 63\% and 66\%.
\end{abstract}
\maketitle

Accurate dielectric and piezoelectric tensors are required to evaluate electrical polarization and electromechanical coupling across functional materials and to screen candidates for electronic and electromechanical applications\cite{Banerjee2022HfO2Review,Moreira2011AlScNFBAR,Petousis2017DielectricScreening,Choudhary2020HighThroughputResponses}. Lattice-mediated (ionic) contributions often provide the largest and most structure-sensitive components of static dielectric and piezoelectric responses\cite{Robertson2004HighDielectricOxides,Ederer2005EpitaxialStrain}. The relevant structural spaces include amorphous solids, defects, interfaces, and configurationally complex crystals, whose representation and sampling demand large or numerous simulation cells\cite{Ceresoli2006AmorphousHfO2,Wang2012AmorphousHighK,Chai2022HZOStackDefects,Hwang2024AlScNIDB,Das2026NanoscaleOrderingAlScN}. DFPT is the established first-principles framework for these response tensors\cite{Gonze1997DFPTResponse,Wu2005DFPTFields}, but obtaining the ionic contribution in an $N$-atom cell requires up to $3N$ force-constant response calculations, each costing $\mathcal{O}(N^3)$, and therefore scales as $\mathcal{O}(N^4)$\cite{Baroni2001DFPTReview}. This computational cost limits routine DFPT calculations for large structural models and extensive materials searches, creating a need for more scalable response-prediction methods\cite{Macheda2024BornLargeCells}.

Existing machine-learning strategies have mainly reduced response-prediction costs through direct regression of macroscopic tensors. Direct tensor models, including ETGNN, GoeCTP, GMTNet, dielectric tensor regressors, and EATGNN-like piezoelectric predictors\cite{Yu2022ETGNN,Hua2024GoeCTP,Hua2026GoeCTPDielectric,Yan2024GMTNet,Mao2024DielectricTensor,Dong2025EATGNNPiezo}, directly predict macroscopic dielectric or piezoelectric tensors from structure at low inference cost. Related crystal-level studies have used learned dielectric properties for screening and descriptor-based structure--property analysis\cite{Morita2020DielectricML,Shimano2023HighDielectricML}, as well as for the discovery of highly anisotropic dielectric crystals\cite{Lou2025AnisotropicDielectric}. However, because each $\mathcal{O}(N^4)$ DFPT calculation contributes only a macroscopic tensor target, these models require large training datasets and make inefficient use of the available response information. Complementary studies have begun to move beyond macroscopic tensor targets by predicting microscopic response quantities, including Born effective charges and real-space response fields\cite{Kutana2025BornEffectiveCharges,MalenfantThuot2024RealSpaceResponse}. Established lattice-dynamical and DFPT formulations express the ionic dielectric tensor as a contraction of Born effective charges with the pseudoinverse of the lattice force-constant matrix\cite{Maradudin1971LatticeDynamics,Gonze1997DFPTResponse,Wu2005DFPTFields}. A recent machine-learning implementation predicts the Born effective charges and obtains the force constants from a pretrained machine-learning force field, then reconstructs the ionic dielectric tensor\cite{Takigawa2025FactorizedIonicDielectric}. By predicting Born effective charges and phonon properties separately and combining them analytically, the factorized model substantially improves ionic dielectric prediction over direct regression; however, the reconstruction remains computationally expensive. The dominant bottleneck is the dense eigendecomposition required to construct the pseudoinverse of the $3N\times3N$ force-constant matrix, whose cost scales as $\mathcal{O}(N^3)$. The preceding finite-difference Hessian construction evaluates an $\mathcal{O}(N)$-cost MLFF on $\mathcal{O}(N)$ displaced configurations and therefore scales as $\mathcal{O}(N^2)$. Its output is also limited to the macroscopic ionic dielectric tensor and therefore does not resolve atom-indexed local contributions to the ionic response.

To predict dielectric and piezoelectric tensors accurately and efficiently, we introduce DART (Direct Atom-Resolved response-Tensor learning) and LARS (Linear-scaling Atom-Resolved Response Solver). DART learns atom-indexed ionic response fields, improving data efficiency, accuracy, and generalization even with small training sets. LARS reconstructs these fields from learned Born effective charges and interatomic potentials with linear scaling throughout the reconstruction workflow, enabling dielectric and piezoelectric predictions for large systems. Both DART and LARS return per-atom ionic dielectric and piezoelectric contributions defined below, enabling local structure--response analysis unavailable from macroscopic tensors.

\begin{figure}[t]
\centering
\includegraphics[width=0.96\columnwidth]{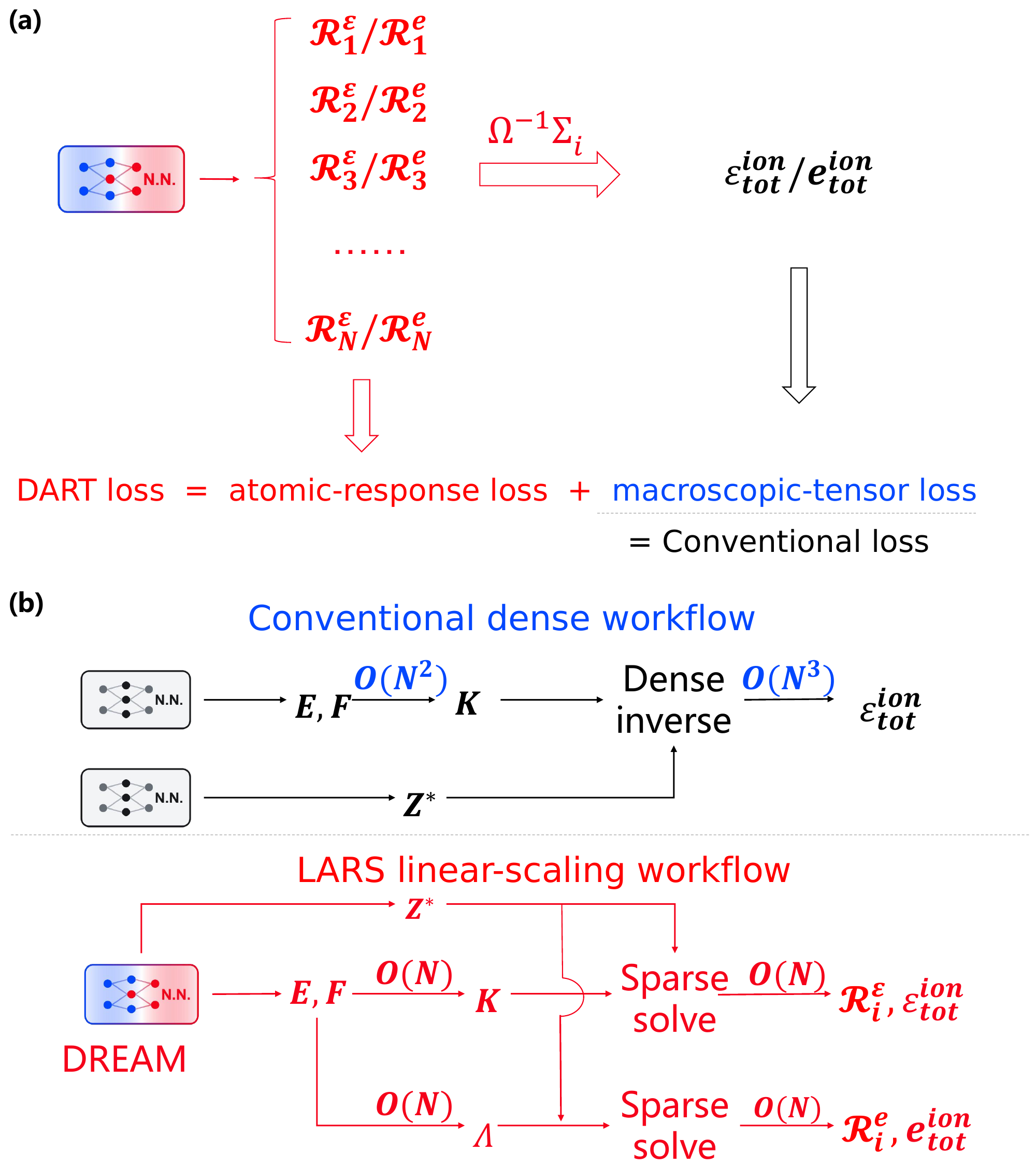}
\caption{Local-response framework. (a) DART learns the DFPT-derived fields $\mathcal{R}^{\epsilon}$ and $\mathcal{R}^{e}$, whose volume averages give the macroscopic dielectric and piezoelectric tensors. (b) LARS reconstructs the same atom-indexed fields from learned Born effective charges, force constants, and internal-strain couplings through sparse applications of $(K^{-1})'$.}
\label{fig:framework}
\end{figure}

\begin{figure*}[t]
\centering
\includegraphics[width=0.95\textwidth]{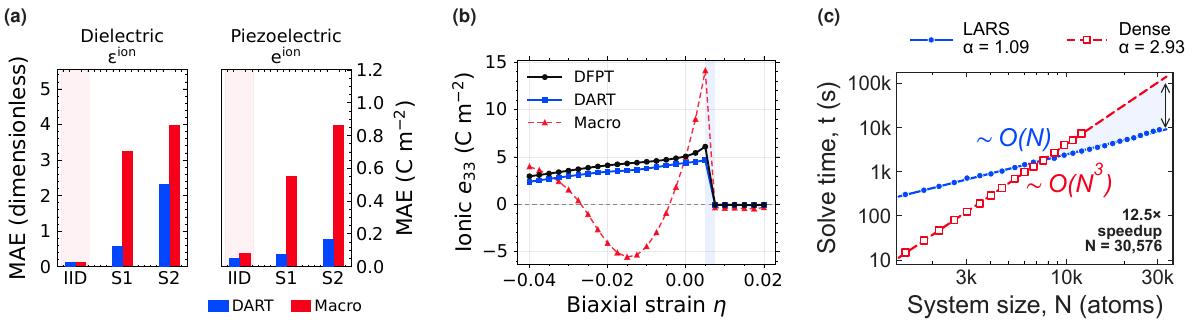}
\caption{Validation of DART and LARS. (a) Structure-level MAEs of DART and direct macroscopic regression for the ionic dielectric and piezoelectric tensors across an independent and identically distributed (IID) test and two out-of-distribution (OOD) schemes: Scheme 1 trains on unstrained disordered alloys and strained ordered configurations and tests on held-out strained-disordered configurations, whereas Scheme 2 extrapolates from $\eta\leq0.02$ to $\eta=0.03$--0.06. Pink shading identifies IID. (b) Ionic $e_{33}$ along a biaxial-strain path in disordered Al$_{0.5}$Sc$_{0.5}$N, predicted using the Scheme 1 models. The shaded region indicates the phase-transition region. (c) Single-CPU sparse-MINRES and dense-pseudoinverse timings for the piezoelectric reconstruction in HZO supercells containing 1,440--30,576 atoms.}
\label{fig:tests}
\end{figure*}

For conciseness, we show only the piezoelectric case; the dielectric case is analogous and is given in the Supplemental Material. The established DFPT expression for the ionic piezoelectric tensor is

\begin{equation}
\begin{aligned}
e^{\rm ion}_{\alpha\mu}
&=\Omega^{-1}C_e\sum_{i,j,\gamma,\delta}
Z^*_{i,\alpha\gamma}(K^{-1})'_{i\gamma,j\delta}\Lambda_{j\delta,\mu}\\
&=\Omega^{-1}\sum_i \mathcal{R}^{e}_{i,\alpha\mu}.
\end{aligned}
\label{eq:dfpt-ionic-responses}
\end{equation}

Here, $i,j$ label atoms, $\alpha,\gamma,\delta$ are Cartesian indices, and $\mu$ labels the strain component. $Z^*_{i,\alpha\gamma}$ denotes the Born effective charge, $K_{i\gamma,j\delta}$ the force-constant matrix, $\Lambda_{j\delta,\mu}$ the internal-strain force coupling, $\Omega$ the cell volume, and $C_e$ a conversion factor. We define $K_{i\gamma,j\delta}=\partial^2U/\partial u_{i\gamma}\partial u_{j\delta}$ and $\Lambda_{i\gamma,\mu}=\partial F_{i\gamma}/\partial\eta_\mu$, where $U$ is the energy, $u$ the atomic displacement, and $F$ the force. The units of $Z^*$, the conversion factor $C_e$, and the Voigt convention are specified in the Supplemental Material. $(K^{-1})'$ denotes the pseudoinverse of $K$. For machine-learning purposes, we define the atomic piezoelectric response in Eq. \eqref{eq:dfpt-ionic-responses} as

\begin{equation}
\begin{aligned}
\mathcal{R}^{e}_{i,\alpha\mu}
&=C_e\sum_{j,\gamma,\delta}
Z^*_{i,\alpha\gamma}(K^{-1})'_{i\gamma,j\delta}\Lambda_{j\delta,\mu}.
\end{aligned}
\label{eq:ipd}
\end{equation}

Thus, both response fields are atom-resolved quantities independent of system size. Because these contributions contain the lattice Green function, they are not strictly local functions of the surrounding geometry. The decomposition nevertheless provides physically well-scaled atom-level labels for neural-network learning.

DART learns these atom-indexed ionic dielectric and piezoelectric response fields from structure. For each training structure, the DFPT-derived local response fields and the corresponding macroscopic tensors serve as the supervision targets. Within DART, a shared DREAM equivariant representation learns the local-response coordinates\cite{Deng2025DreamAllegro,Yu2025DreamFerroelectricSi}. Task-specific tensor readouts map this representation to the local ionic fields and the corresponding electronic or clamped-ion responses. With $\mathbf h_i$ denoting the learned feature of atom $i$ and $\theta$ the trainable model parameters, the piezoelectric branch predicts the local field and its macroscopic volume average:

\begin{equation}
\widehat{\mathcal R}^{e}_{i,\alpha\mu}
=f^{e}_{\theta,\alpha\mu}(\mathbf h_i),
\quad\widehat e^{\rm ion}_{\alpha\mu}
=\Omega^{-1}\sum_i\widehat{\mathcal R}^{e}_{i,\alpha\mu}.
\label{eq:dart-learning}
\end{equation}

Besides, the loss jointly supervises the macroscopic ionic piezoelectric tensor and the atomic response tensors, as illustrated in Fig. \ref{fig:framework}(a):

\begin{equation}
\mathcal L^{e}_{\rm ion}
=\lambda_e\operatorname{MSE}(\widehat e^{\rm ion},e^{\rm ion})
+\frac{\lambda_R}{N}\sum_{i=1}^{N}\operatorname{MSE}(\widehat{\mathcal R}^{e}_i,\mathcal R^{e}_i).
\label{eq:dart-loss}
\end{equation}
The loss is written for a structure with $N$ atoms, with MSE averaging over tensor components; $\lambda_e$ and $\lambda_R$ weight the macroscopic and atomic terms, respectively. Pooled system-level features separately predict the electronic dielectric and clamped-ion piezoelectric tensors.

Trained on DFPT response labels, DART bypasses the inverse-Hessian solve at inference and directly predicts the macroscopic response together with its atom-indexed local response field. We evaluate DART against direct macroscopic regression under an independent and identically distributed (IID) test and two out-of-distribution (OOD) schemes: Scheme 1 trains on unstrained disordered alloys and strained ordered 1AlN/1ScN configurations, reserving the complete strained-disordered trajectories for testing, and Scheme 2 extrapolates from $\eta\leq0.02$ to $\eta=0.03$--0.06 (Fig. \ref{fig:tests}(a)). Under Scheme 1, $\mathcal{R}^{e}$ supervision reduces the ionic piezoelectric MAE from 0.5539 to 0.0741 C/m$^2$ and improves $R^2$ from $-3.1938$ to 0.9511. Figure \ref{fig:tests}(a) shows that DART substantially improves the accuracy of ionic dielectric and piezoelectric predictions over direct macroscopic regression in the OOD tests. Additional results are provided in the Supplemental Material.

Using the DART model trained under Scheme 1, we predict the strain-dependent response of disordered Al$_{0.5}$Sc$_{0.5}$N and compare it with direct macroscopic regression. Along a representative path, DFPT identifies a polar branch under compression and weak tension, along which ionic $e_{33}$ rises as biaxial tension approaches the transition and collapses near $\eta\simeq0.0075$. DART tracks both the rise and the collapse. The direct tensor regressor instead predicts the wrong trend and sign in part of the compressed region and exaggerates the pretransition spike (Fig. \ref{fig:tests}(b)). This extrapolation test shows why local-response supervision is useful beyond interpolation.

Whereas DART learns the local response fields directly, LARS reconstructs the same atom-indexed fields from learned microscopic ingredients without $\mathcal{O}(N^4)$ DFPT response labels. The sparse reconstruction is performed for the finite-range learned potential trained with DREAM. The DREAM model supplies the Born effective charges, force constants, and internal-strain couplings used by LARS\cite{Deng2025DreamAllegro,Yu2025DreamFerroelectricSi}. The piezoelectric form in Eq. \eqref{eq:dfpt-ionic-responses}, together with the dielectric counterpart in the Supplemental Material, shows that each reconstruction applies the force-constant pseudoinverse to its own right-hand side: $Z^*$ for the dielectric response and $\Lambda$ for the piezoelectric response. For either reconstruction, collect the perturbation directions in a matrix $B$, with $B_{i\gamma,\beta}=Z^*_{i,\beta\gamma}$ for the dielectric response and $B_{i\gamma,\mu}=\Lambda_{i\gamma,\mu}$ for the piezoelectric response, and define the response matrix

\begin{equation}
X_B\equiv (K^{-1})'B.
\label{eq:lars-response}
\end{equation}

The acoustic sum rules ensure zero net driving force for both right-hand sides; residuals in the learned quantities are treated in the Supplemental Material. LARS obtains $X_B$ by solving the linear system:

\begin{equation}
KX_B=B.
\label{eq:lars-force-balance}
\end{equation}

We fix the arbitrary rigid translation by requiring the atomic displacements in each column of $X_B$ to sum to zero. Substituting $X_B$ into the appropriate DFPT expression gives the atomic polarizability or atomic piezoelectric response field defined above. For the piezoelectric response ($B=\Lambda$),

\begin{equation}
\mathcal R^e_{i,\alpha\mu}
=C_e\sum_\gamma Z^*_{i,\alpha\gamma}(X_B)_{i\gamma,\mu}.
\label{eq:lars-atomic-piezo}
\end{equation}

As shown in Fig. \ref{fig:framework}(b), conventional reconstruction forms the pseudoinverse of the dense $3N\times3N$ force-constant matrix at $\mathcal{O}(N^3)$ cost. LARS instead uses preconditioned MINRES to solve $KX_B=B$; finite-range interactions give $K$ only $\mathcal{O}(N)$ nonzero entries, so each iteration costs $\mathcal{O}(N)$ and the solve scales linearly when the iteration count is independent of $N$. The local-block decomposition (see Supplemental Material) enables parallel sparse operations across multiple CPUs. The same locality also makes construction of the force-constant matrix $K$ linear in system size [Fig. \ref{fig:framework}(b)]. Conventional finite differences cost $\mathcal{O}(N^2)$, requiring $\mathcal{O}(N)$ full-system force evaluations at $\mathcal{O}(N)$ cost each. We achieve $\mathcal{O}(N)$ construction by modifying DREAM to evaluate $N$ local Hessian blocks of bounded size in parallel and assemble the global sparse Hessian.

% MD-COMMENT:  Superseded caption retained only in the original revision record.
\begin{figure*}[t]
\centering
\includegraphics[width=0.94\textwidth]{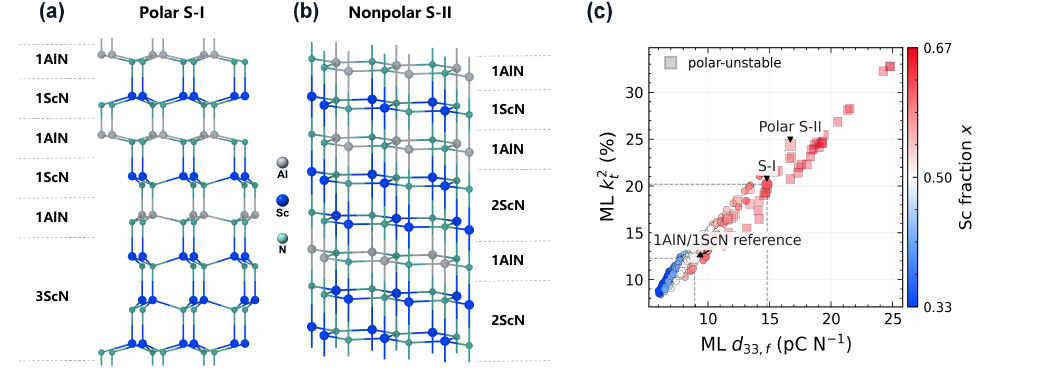}
\caption{DART screening of AlScN superlattices. (a) Ball-and-stick side view of the top verified candidate S-I. (b) Nonpolar relaxed state obtained from the $1\mathrm{AlN}/1\mathrm{ScN}/1\mathrm{AlN}/2\mathrm{ScN}/1\mathrm{AlN}/2\mathrm{ScN}$ stacking at the same average composition $x=0.625$, compared with S-I in panel (a). (c) ML ranking of the 314 zero-strain AlN/ScN superlattices by the laterally clamped longitudinal piezoelectric strain coefficient $d_{33,f}$ and the thickness-mode electromechanical coupling coefficient $k_t^2$. The ordered 1AlN/1ScN superlattice is marked as the reference; the DFPT-verified responses are reported in the Supplemental Material.}
\label{fig:dart-application}
\end{figure*}

For HZO supercells containing 1,440--30,576 atoms, the LARS piezoelectric solve time grows nearly linearly with system size [Fig. \ref{fig:tests}(c)]. At $N=30{,}576$, the measured LARS solve is 12.5 times faster than the dense-method fit extrapolation. We further test LARS on 768-atom amorphous HfO$_2$ supercells, linking local coordination environments to atomic polarizability and providing a microscopic explanation for the enhanced ionic dielectric response (see Supplemental Material).

DART directly yields the dielectric and piezoelectric tensors in a single inference pass, whereas LARS retains the reconstruction cost and soft-mode error amplification inherent to reconstruction methods. When DFPT dielectric and piezoelectric response data are available, DART therefore usually provides faster and more accurate predictions; LARS, however, requires no $\mathcal{O}(N^4)$ DFPT response labels from dataset preparation through inference and retains linear scaling for large systems.

We next apply DART to screen AlScN superlattices. The ferroelectricity and polarization switching of AlScN\cite{Fichtner2019AlScNFerroelectric,Wang2021PiezoelectricSwitchingAlScN} have attracted considerable attention, and its piezoelectric response has also been extensively studied\cite{Akiyama2009ScAlNPiezo,Tasnadi2010AnomalousPiezo}. In AlScN, Sc substitution in wurtzite AlN enhances the piezoelectric response through structural softening and internal-strain sensitivity, while residual stress provides an additional modulation\cite{Osterlund2021AlScNResidualStress}. This electromechanical coupling makes AlScN suitable for acoustic resonators and surface-acoustic-wave devices\cite{Moreira2011AlScNFBAR,Wang2014AlScNSAW,Park2019ScAlNFBAR,Wang2020AlScNFBAR}. We exhaustively enumerate 314 periodic, symmetry-unique binary AlN/ScN sequences along the c axis, spanning the Sc fraction $x$ in $\mathrm{Al}_{1-x}\mathrm{Sc}_x\mathrm{N}$ over $x\in[1/3,2/3]$ and periods of 4--12 layers, with no cyclic AlN or ScN block longer than four layers (Fig. \ref{fig:dart-application}(c)). The search covers structures with split ScN blocks, moderately thick ScN slabs, and high interface densities. We rank the zero-strain polar candidates by $d_{33,f}=e_{33}/C_{33}^{E}$ and the thickness-mode electromechanical coupling coefficient $k_t^2$ against the ordered one-layer AlN/one-layer ScN reference, using $C_{33}^{E}$ from DFT calculations. The in-plane and shear strains are clamped, with zero normal stress; superscript $E$ denotes fixed electric field. The dielectric and piezoelectric training sets contain no complex superlattices; predictions for all 314 candidates are therefore structure-level extrapolations. For six DFPT-rechecked superlattices, the dielectric and piezoelectric MAEs are 1.07 in the relative permittivity $\epsilon_{33}$ and 0.133 C/m$^2$ in $e_{33}$, respectively. Candidates are verified with DFT/DFPT response calculations in VASP\cite{Kresse1996VASP} and finite-strain $C_{33}^{E}$. We evaluate $k_t^2$ from the calculated material tensors under the thickness-mode convention,

\begin{equation}
k_t^2=\frac{e_{33}^2}{\epsilon_0\epsilon_{33}^{\eta}C_{33}^{E}+e_{33}^2},
\label{eq:thickness-coupling}
\end{equation}

where $e_{33}=e_{33}^{\mathrm{el}}+e_{33}^{\mathrm{ion}}$ is the total proper piezoelectric coefficient and $\epsilon_{33}^{\eta}=\epsilon_{33}^{\mathrm{el}}+\epsilon_{33}^{\mathrm{ion}}$ is the total relative permittivity at fixed strain; $\epsilon_0$ is the vacuum permittivity. It is used as a screening metric rather than a device-level effective coupling coefficient.

Under this convention, the 1AlN/1ScN reference gives $d_{33,f}=8.9239$ pC/N and $k_t^2\simeq12.29\%$. DFPT gives $d_{33,f}=14.511$ pC/N and $k_t^2=20.46\%$ for S-I, compared with 8.9239 pC/N and 12.29\%, respectively, for the 1AlN/1ScN reference. Figure \ref{fig:dart-application}(a,c) shows the S-I structure and the screening ranking. Other stacking sequences in the broader superlattice screen also show high response. Increasing Sc content strengthens the competition between the polar wurtzite-derived state of AlScN and other structural minima\cite{Tasnadi2010AnomalousPiezo}. At $x=0.625$, the selected $1\mathrm{AlN}/1\mathrm{ScN}/1\mathrm{AlN}/1\mathrm{ScN}/1\mathrm{AlN}/3\mathrm{ScN}$ stacking yields the polar S-I candidate, whereas the comparison stacking shown in Fig. \ref{fig:dart-application}(b) relaxes to a nonpolar state with $e_{33}\simeq0$ and is therefore excluded from the piezoelectric ranking. For the S-I cation sequence, the polar structure lies 2.14 meV/atom below the tested $c$-axis-collapsed configuration. Phonon calculations support its dynamical stability, with the near-$\Gamma$ convergence limits documented in the Supplemental Material. In 100 ps NVT molecular-dynamics simulations with a fine-tuned DPA-4 potential\cite{Li2026DPA4}, 512-atom S-I supercells remain structurally stable at 300, 600, 900, and 1200 K: the Al/Sc layer sequence is preserved, the drift-corrected mean-square displacement stays below 0.10 \AA$^2$, and no N--N dimer is detected in the saved trajectories. The enhanced $d_{33,f}$ and $k_t^2$ of S-I point to layer ordering as a route to tune electromechanical response.

In summary, we develop an atom-resolved framework for ionic dielectric and piezoelectric response, in which atomic polarizability and atomic piezoelectric response fields are obtained either by direct learning in DART or by sparse reconstruction in LARS. DART learns atom-indexed responses for extrapolative screening, whereas LARS reconstructs them from learned microscopic ingredients with linear scaling and without $\mathcal{O}(N^4)$ DFPT response labels. By learning atom-indexed responses, DART makes high-throughput screening of complex AlN/ScN stacking spaces possible and shows that layer ordering tunes piezoelectric response. We identify S-I, an AlN/ScN superlattice that retains a stable polar structure at a high Sc fraction of 0.625, with $d_{33,f}$ and $k_t^2$ exceeding those of the ordered 1AlN/1ScN reference by 63\% and 66\%, respectively. This framework provides a general basis for atom-resolved ionic dielectric and piezoelectric response calculations across structurally complex materials spaces, supporting high-throughput screening, physical interpretation, and large-system simulations.

\textit{Acknowledgments.---} We acknowledge financial support from the National Key R\&D Program of China (Grant No. 2022YFA1402901), the National Natural Science Foundation of China (NSFC, Grant No. 12188101), the Shanghai Science and Technology Program (No. 23JC1400900), the Guangdong Major Project of Basic and Applied Basic Research (Future Functional Materials under Extreme Conditions, Grant No. 2021B0301030005), the Shanghai Pilot Program for Basic Research at Fudan University (No. 23TQ017), the robotic AI-Scientist platform of the Chinese Academy of Sciences, and the New Cornerstone Science Foundation.

\end{document}